\pdfoutput=1
\documentclass[showpacs,prb,twocolumn]{revtex4}
\usepackage{amsmath}
\usepackage{amssymb}
\usepackage{color}
\usepackage{graphics}
\usepackage{graphicx}
\usepackage{pdfsync}

\def\boldit#1{\mbox{\boldmath$#1$}}

\def\q{\quad}

\begin{document}

\title {Oblique launching of optical surface waves by a subwavelength slit}

\author{ A.~Yu.~Nikitin$^{1,2}$}
\email{alexeynik@rambler.ru}
\author{ F.~J.~Garc\'{i}a-Vidal$^3$}
\author{ L.~Mart\'{i}n-Moreno$^1$}
\email{lmm@unizar.es}
 \affiliation{$^1$ Instituto de Ciencia de Materiales de Arag\'{o}n and Departamento de F\'{i}sica de la Materia Condensada,
CSIC-Universidad de Zaragoza, E-50009, Zaragoza, Spain \\
$^2$A.Ya. Usikov Institute for Radiophysics and Electronics, Ukrainian Academy of
Sciences, 12 Acad. Proskura Str., 61085 Kharkov, Ukraine\\
$^3$ Departamento de F\'{i}sica Te\'{o}rica de la Materia Condensada, Universidad Aut\'{o}noma de Madrid, E-28049
Madrid, Spain}

\begin{abstract}
The electromagnetic field on the metal surface launched by a subwavelength slit is analytically studied, for the case when the fundamental mode inside the slit has a wavevector component along the slit axis (conical mount). Both near-field and far-field regions are discussed, and the role of surface plasmon-polaritons and Norton waves is revealed. It is shown that the distance from the slit at which Norton waves are more intense than surface plasmons decrease with parallel wavevector. Additionally, it is found that the $s$-polarization component, while present for any non-zero parallel wavevector, only weakly contributes to the NWs.
\end{abstract}

 \maketitle

\section{Introduction}

Launching surface plasmon-polaritons (SPP) along metal surfaces has recently attracted a lot of interest for its
possible application in integrated optical
devices.\cite{EbbesenNature03,Maradudin05,Novotny,Maier} One of the most common configurations used employs systems with translational symmetry in one direction, as a subwavelength slit\cite{LalanneNature06,FLTNature07} or a line defect\cite{KrennOL07,QuidantOE10}.

Very recently, several studies have been devoted to finding simple analytical models for the electromagnetic (EM) field radiated by a single slit, which provides insight into the relevant physical processes.\cite{UngShengOE08,LalanneSciRep09,SoukoulisPRB09,NikitinNJP09,NikitinPSS10} It has been found that the field at the surface presents a rich behavior as a function of both frequency and distance to the slit. Typically, at distances less that 2-3 wavelengths, the field presents a complex spatial dependence\cite{FLTPRB05} (which is sometimes phenomenologically described as composed of a SPP plus a ``creeping" or ``quasi-cilyndrical" wave, CW \cite{LalanneNature06,LalanneSciRep09}). At larger distances, there is an intermediate regime where the EM field is dominated by the SPP contribution, and a long-distance regime where the SPP has become negligible and the field is that of a Norton wave (NW). All the previously cited works focused on the case of field propagation perpendicular to the slit, and did not addressed the situation of non-normal incidence of light into the defect. This is a serious deficiency as SPP scattering effects are easier to detect when the SPP is launched obliquely, i.e., when its wavevector has a component along the slit axis. \cite{Drezet_thinEPL06,GordonPRB06,SanchezGilAPL07,NikitinPRB07,KrennOL07,QuidantOE10}

\begin{figure}[thb!]
\includegraphics[width=8cm]{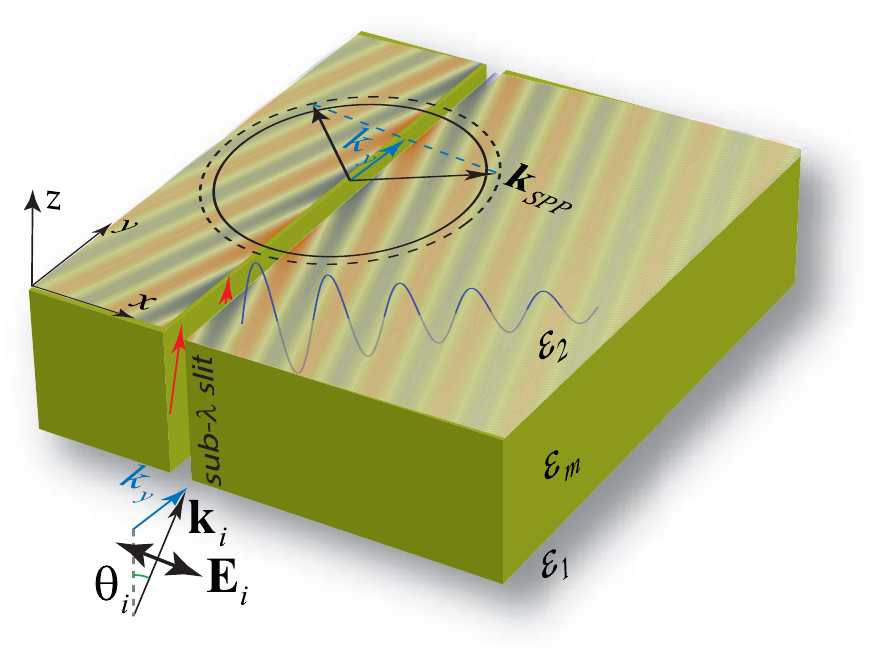}
\caption{(color online). The geometry of the studied system. A plane electromagnetic wave impinges onto a subwavelength slit, placed in an optically thick film. The angle of incidence is arbitrary provided the electric field points perpendicular to the slit axis. The slit aperture at the transmission region generates a field at the surfaces, propagating non-perpendicularly to the slit axis. The solid circle represents a slice of the light cone for a fixed frequency, while the dashed one is the slice of the ``SPP cone''. Arrows inside the slit indicate the propagation direction of the fundamental mode.}\label{Fig1}
\end{figure}

In this paper we fill this gap, presenting an analysis of the EM fields on the metal surface obliquely launched by a subwavelength one-dimensional aperture. Such a launching could be realized by illuminating a subwavelength slit in a thick metal film by a plane wave in the conical mount, where the incident wavevector and the electric field have a non-zero components along the slit axis and across the slit, respectively (see Fig.~\ref{Fig1}). From now on, we will denote this configuration as ``oblique incidence". The subwavelength slit ``filters'' the EM field inside it, so that only a capacitor-like fundamental mode transfers the energy to the outgoing face of the film, all other modes inside the slit being exponentially suppressed. For oblique incidence, the $k$-vector of the fundamental mode has a finite projection along the slit axis and therefore the field emerging from the outgoing aperture gains both $p$- and $s$-polarization components. Here we do not consider the part of the problem
related to the transmission and reflection efficiencies, but concentrate on the electric field pattern on the outgoing face of the film. We thus only need to know the Green's function of our system and the field at the outgoing face of the slit. For this we apply the mathematical methods described in Ref.~\onlinecite{FelsenMarcuvitz}, which have been previously used for obtaining the field at the metal surface radiated both by a single slit for $k_y=0$\cite{UngShengOE08,NikitinNJP09,NikitinPSS10} and by a subwavelength hole\cite{NikitinPRL2011}.

Anticipating things, we would like to stress that the situation where $k_y\neq0$, presents two main differences with the case $k_y=0$: (\emph{i}) the transmitted electric fields have components parallel to the slit axis and (\emph{ii}) the propagation length of the excited SPP along the direction across the slit diminishes approximately as $L_{SPP}\propto\cos\theta_i$. The latter fact favors bringing closer to the slit the region where Norton waves dominate.

\section{Analytical solution}

Let us consider a plane monochromatic wave incident onto a thick metallic film with a subwavelength slit centered at $X=0$. The wavevector of the incident wave is $\textbf{k}_i$, forming an angle $\theta_i$ with respect to the Oy axis, and the wave is polarized perpendicularly to the slit (see Fig.~\ref{Fig1}). If the incoming dielectric media has the dielectric permittivity $\varepsilon_1$, $k_y = \sqrt{\varepsilon_1}k_\omega\sin\theta_i$, where $k_\omega=2\pi/\lambda$. Due to the translational symmetry of the system, the $y$-component of the wavevector is conserved in the transmission process.
This places constrains onto the angular interval where transmitted radiation can be found. If the outgoing half-space has dielectric permittivity $\varepsilon_2$, the range of the allowable angles is given by $k_y<k_\omega\sqrt{\varepsilon_2}$, or $\sin\theta_i<\sqrt{\varepsilon_2/\varepsilon_1}$ (otherwise, the field has evanescent character in the outgoing region).
For example, if the field is incident from the glass substrate with $\varepsilon_1=2.25$ and the outgoing medium is vacuum, the permissible angles are $\theta_i< 41.81^\circ$.

\subsection{General analytical expression for the field}

Let us start with a note on the notation used: throughout the paper, distances in lower case letters are expressed in dimensionless units, defined as $x = k_\omega X$, $z = k_\omega Z$, and the dimensionless wavevector components are denoted as $q_{x,y}=k_{x,y}/k_\omega$ (so that, for instance, the light cone in vacuum corresponds to $q=1$).

Starting from Lippmann-Schwinger integral equation and taking into account that the field in the slit points along the  $x$-direction $\mathbf{E}(x,y,z) \simeq \mathbf{e}_x E_x(x,z) e^{i q_y y}$,
the expression for the transmitted field simplifies to (see Appendix for details)
\begin{equation}\label{e1}
\begin{split}
\mathbf{E}(x,z) = C\,\int_L dx'\mathbf{G}(x-x',z)E_x(x',z=-\delta),
\end{split}
\end{equation}
where $C = i\sqrt{\varepsilon_m-\varepsilon}/k_\omega$. The dielectric constants
of dielectric in the outgoing region and the metal are $\varepsilon\equiv\varepsilon_2$ and $\varepsilon_m$, respectively, and $\delta$ is the skin depth in the metal. The integration in $x'$ is performed across the slit area.
The cyclic dependency upon the coordinate $y$, $\propto e^{iq_y y}$ is omitted here and in what follows. $\mathbf{G}(x,z)$ is the $x$-column of the Green's Dyadic, $\mathbf{G}\equiv\hat{G}\mathbf{e}_x$ (with $\mathbf{e}_x$ being unitary vector along $Ox$ axis), whose angular representation for the case of arbitrary $k_y$ reads
\begin{equation}\label{e2}
\begin{split}
\mathbf{G}^p(x,z) = \frac{ik_\omega}{4\pi}\int \frac{dq_x}{q^2\sqrt{\varepsilon_m\varepsilon}}t^p
\begin{pmatrix}
q^2_xq_z\\
q_xq_yq_z\\
-q_xq^2
\end{pmatrix}e^{i(q_xx+q_{z}z)},\\
\mathbf{G}^s(x,z) = \frac{ik_\omega}{4\pi}\int \frac{dq_x}{q^2q_{zm}} t^s
\begin{pmatrix}
q^2_y\\
-q_xq_y\\
0
\end{pmatrix}e^{i(q_xx+q_{z}z)},
\end{split}
\end{equation}
where $\mathbf{G} = \mathbf{G}^p + \mathbf{G}^s$, indices ``$p$'' and ``$s$'' stay for corresponding polarizations and $t_p$ and $t_s$ are Fresnel transmission coefficients for the metal-dielectric interface, given by
\begin{equation}\label{e3}
\begin{split}
t^s = \frac{2q_{zm}}{q_{zm}+q_{z}},\q t^p = \sqrt{\frac{\varepsilon_m}{\varepsilon}}\frac{2q_{zm}\varepsilon}{q_{z}\varepsilon_m + q_{zm}\varepsilon},
\end{split}
\end{equation}
with $q^2 = q_x^2 + q_y^2$, $q_z = \sqrt{\varepsilon-q^2}$, $q_{zm} = \sqrt{\varepsilon_m-q^2}$. The branches of $q_z$, $q_{zm}$ must be chosen in accordance with the radiation conditions $\mathrm{Im}(q_z,q_{zm})\geq0$. It should be noted that for $q_y=0$ the integrals transform to the case of in-plane launching (see Refs.\onlinecite{NikitinNJP09,NikitinPSS10}) and $s$-components of the fields vanish.

For narrow (subwavelength) slits, the field inside the slit can be taken as independent of $x$ and, therefore, the outgoing slit aperture is equivalent to the effective two-dimensional electric dipole located on the metal surface\cite{NikitinNJP09}: $\mathbf{E}(x,z) = \mathbf{G}(x,z)p^{eff}$, where $p^{eff} = aCE_x(0,z=-\delta)$ and $a$ is the width of the slit (in dimensionless units).

The numerical computation of the integrals in Eq.~\eqref{e2} are notoriously difficult,
due to the simultaneous presence of poles, branch cuts, and strongly oscillatory factors. Using a special mathematical treatment based on the steepest descent method,\cite{FelsenMarcuvitz} an accurate analytical representation of the asymptotic behavior of the field is possible.
We present here the final result for the field at the metal surface $z=0$; the mathematical details can be found in the Appendix. Introducing the following notation
\begin{equation}\label{e4}
\begin{split}
\mathbf{G}(x,z=0) = \frac{ik_\omega}{2\pi}\mathbf{g}(x),
\end{split}
\end{equation}
we have  in the region $q_0x\gg1$:
\begin{equation}\label{e5}
\begin{split}
&\mathbf{g}(x) \approx i\pi \mathbf{C}_p e^{iq_{xp}x}\mathrm{erfc}(- is_p\sqrt{q_0x})+ e^{iq_0x}\sqrt{\frac{\pi}{q_0x}}\frac{\mathbf{C}_p}{s_p} + \\
& \frac{\sqrt{\pi}e^{iq_0x}}{4q_0x\sqrt{q_0x}}\left[\frac{2\mathbf{C}_p }{s_p^3} + 2\sqrt{2}e^{-i\frac{3\pi}{4}}(\mathbf{f}^s+\mathbf{f}^p)\right].
\end{split}
\end{equation}
In this equation $q_0=\sqrt{\varepsilon-q_y^2}$ presents the inverse spatial period along the $x$-axis of the algebraically-decaying terms, $q_p = \sqrt{\varepsilon\varepsilon_m/(\varepsilon+\varepsilon_m)}$ is the modulus of the in-plane component of SPP momentum and $q_{xp} = \sqrt{q_p^2-q_y^2}$ is its $x$-component. The factor $s_p= e^{-i\pi/4}\sqrt{q_{xp}/q_0-1}$, appearing in the argument of the complementary error function, $\mathrm{erfc}$, is the position of the pole in the complex plane where the steepest-descent integration is made (see Appendix). It has an important significance, being responsible for the asymptotic expansion of the error function, whose argument is the the square root of the so called numerical distance introduced by Sommerfeld. Another property of $s_p$ is that $|s_p|^2$ quantify the distance in complex $q-$space between the SPP pole and the branch-point placed at $q=\sqrt{\varepsilon}$ (i.e. at $q_z=0$).

The terms in Eq.~\eqref{e5} containing $\mathbf{C}_p$ come from the singular part of the integrals in Eq.~\eqref{e2} appearing for $p$-polarization. $\mathbf{C}_p$ are the residues at $q_x=q_{px}$:
\begin{equation}\label{e6}
\begin{split}
\mathbf{C}_p =  \frac{\varepsilon_mq_{zp}\sqrt{\varepsilon_m-\varepsilon}}{q_p^2\left(\varepsilon^2-\varepsilon^2_m\right)}
\begin{pmatrix}
q_{xp}q_{zp}\\
q_{y}q_{zp}\\
-q_p^2
\end{pmatrix},
\end{split}
\end{equation}
where $q_{zp} = \varepsilon/\sqrt{\varepsilon+\varepsilon_m}$.

Finally, $\mathbf{f}^{s,p}$ are contributions to the algebraically-decaying term, that in this 1D geometry go like $\sim1/x^{3/2}$, which dominate the far-field region at the surface (unless the system is completely absorptionless)
\begin{equation}\label{e7}
\begin{split}
\mathbf{f}^s =  \frac{2q_{y}q_0^2}{\varepsilon(\varepsilon_m-\varepsilon)}
\begin{pmatrix}
-q_{y}\\
q_0\\
0
\end{pmatrix}, \\
\mathbf{f}^p =  \frac{2q_0^3}{\varepsilon^2}
\begin{pmatrix}
q_0\\
q_{y}\\
\frac{\varepsilon_m}{\sqrt{\varepsilon_m-\varepsilon}}
\end{pmatrix}.
\end{split}
\end{equation}

\begin{figure}[thb!]
\includegraphics[width=7cm]{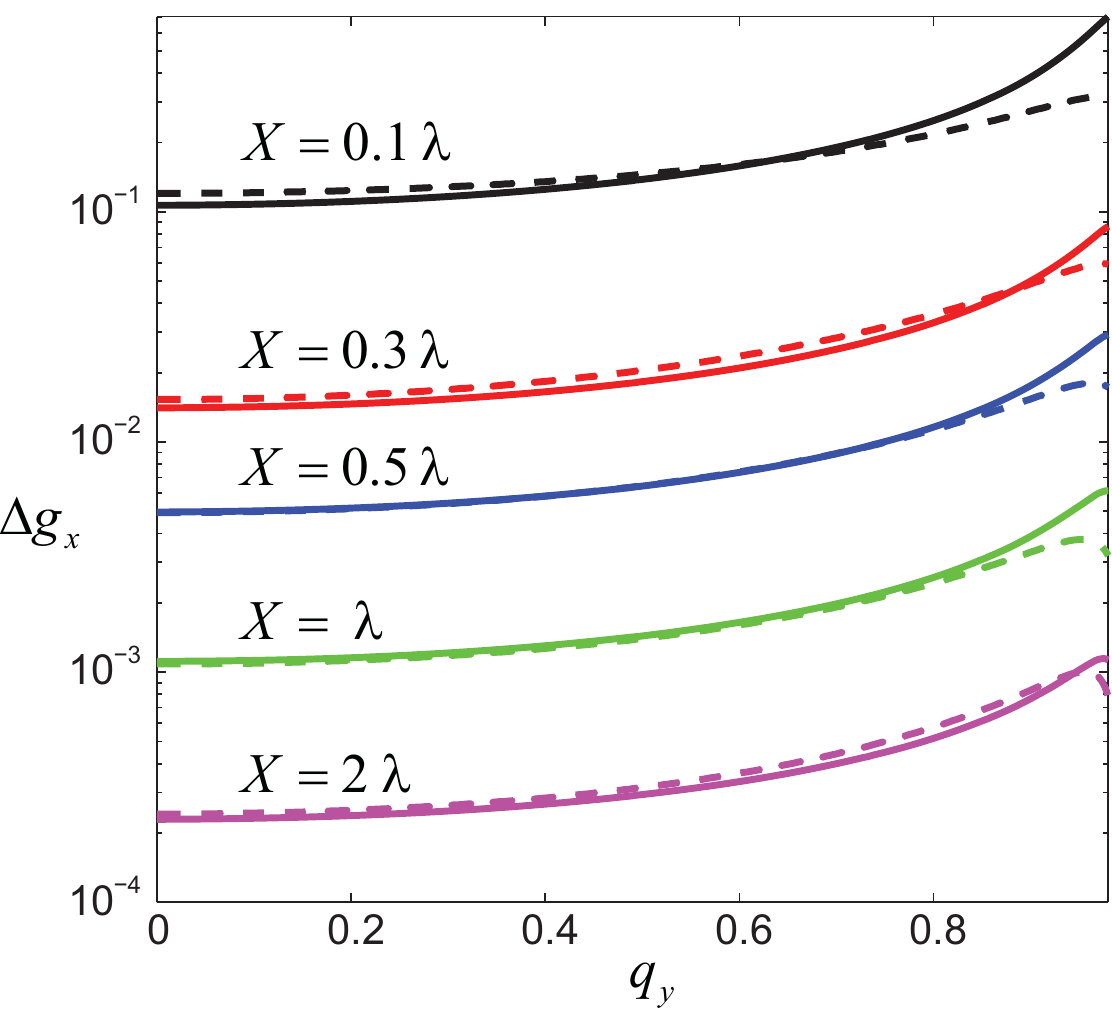}
\caption{(color online). The relative error $\Delta g_x$ as a function of $q_y$ in the case of a gold surface. Continuous curves correspond to $\lambda=700 nm$, while discontinuous ones are for $\lambda= 540 nm$.}\label{Fig2}
\end{figure}

It is remarkable that the region for validity of the solution \eqref{e4}-\eqref{e7} is much less restrictive than $q_0x\gg1$ (i.e. $x\gg 1/\sqrt{\varepsilon-q_y^2}$), similarly to what occurred in the case of $q_y=0$\cite{NikitinNJP09}. Notice, nevertheless, that as
$q_y$ increases this formal condition is fulfilled for larger values of $x$, which explains that, at a fixed distance, the relative error in the field increases with $q_y$.
To characterize the relative error, we have introduced the following function $\Delta g_\alpha = \left|(g_\alpha-g^{num}_\alpha)/g^{num}_\alpha\right|$ with $\alpha=x,y,z$, where $g^{num}_\alpha$ correspond to precise numeric calculations and $g_\alpha$ is given by Eqs.~\eqref{e5}-\eqref{e7}. Except for $q_y \approx 1$
this error is not very sensitive to the value of $\epsilon_m$, and, therefore, almost independent of wavelength (from the optical region to longer wavelengths).
We have checked that the relative error does not exceed a few percents for distances as small as $X\sim\lambda/10$, and already for $X$ of order of a wavelength the error has reduced down to 0.1\% at $q_y=0$ (and $\sim 0.5\%$ at $q_y \approx 1$), see Fig.~\ref{Fig2}.

\subsection{Perfect electric conductor limit}

Before analyzing the case of a real conductor, let us consider the limiting case of a
perfect electric conductor (PEC), characterized by $\varepsilon_m\rightarrow -\infty$.
Then the electric field at the surface is perpendicular to it and, for a very thin slit, can be analytically calculated using directly Eqs.~\eqref{e1},\eqref{e2}:
\begin{equation}\label{e8}
\begin{split}
\mathbf{E}^{PEC}(x,0) = -\mathbf{e}_z\frac{q_0}{2} aH^{(1)}_1(q_0|x|)E_x^{PEC}(0,0),
\end{split}
\end{equation}
with $H^{(1)}_1$ being Hankel function of first order.

The PEC limit of the asymptotic expansion has to be taken with care, as the Green's Dyadic tends to zero as $\mathbf{G}\sim1/\sqrt{\varepsilon_m}$. This is natural, since the dipole placed on the metal interface and oriented along it, cannot radiate due to cancellation of the field by the image dipole. However, the effective dipole of the slit diverges as $p^{eff}\sim\sqrt{\varepsilon_m}$, so that the product $\mathbf{G} \, p^{eff}$ remains finite. If we substitute Eqs.~\eqref{e4}-\eqref{e7} into Eq.~\eqref{e1} and perform the PEC limit, we arrive at the asymptotic expansion for Eq.~\eqref{e8}. This expansion consists of just one term, where the Hankel function is replaced by its asymptotic term $H^{(1)}_1(q_0|x|)=-\sqrt{2i/(\pi q_0x)}e^{iq_0x}$.
Thus, our asymptotic expansion recovers the PEC result, up to terms of order $O(x^{-5/2})$.

\subsection{Far-field asymptotic}\label{Sff}

In case of long distances, or, more precisely, when $xq_0|s_p|\gg1$, we can obtain a simplified expression from Eq.~\eqref{e5} (exact up to $O(x^{-5/2})$):
\begin{equation}\label{e9}
\begin{split}
\mathbf{g}(x) = \mathbf{g}_{SPP}(x) + \mathbf{g}_{NW}(x),
\end{split}
\end{equation}
where $\mathbf{g}_{SPP}$ is the contribution from the SPP pole (arising from the first term in the asymptotic expansion of the complementary error function)
\begin{equation}\label{e10}
\begin{split}
\mathbf{g}_{SPP}(x) = 2\pi i \mathbf{C}_p e^{iq_{xp}x},
\end{split}
\end{equation}
and $\mathbf{g}_{NW}(x) = \mathbf{g}_{NW}^p(x)+\mathbf{g}_{NW}^s(x)$ is an algebraically-decaying term, with contribution from both polarizations
\begin{equation}\label{e11}
\begin{split}
\mathbf{g}_{NW}^\sigma(x) = \frac{\sqrt{\pi}e^{-i\frac{3\pi}{4}}}{q_0\sqrt{2q_0}}\mathbf{f}^\sigma\frac{e^{iq_0x}}{x\sqrt{x}} , \q \sigma=p,s.
\end{split}
\end{equation}

This field component presents the two-dimensional analog of the Norton waves (NW) discovered theoretically almost a century ago by Norton\cite{Norton}, when analyzing the radiation of radio waves by a point dipole placed over the Earth surface (represented by a lossy dielectric).

We would like to stress that the approximation \eqref{e9}-\eqref{e11} is not applicable for PECs since, in this case, $s_p=0$ and the condition for validity of the asymptotic expansion is not fulfilled for any distance.

In the remaining part of the article we analyze the dependency of the fields upon $q_y$. We concentrate on the case of a vacuum-gold interface setting $\varepsilon=1$, $\varepsilon_m=\varepsilon_{Au}$. The effects generated by changing $\varepsilon$ were considered in Ref.~\onlinecite{NikitinPSS10}.

\section{Illustrative results and their discussion}

\begin{figure}[thb!]
\includegraphics[width=8cm]{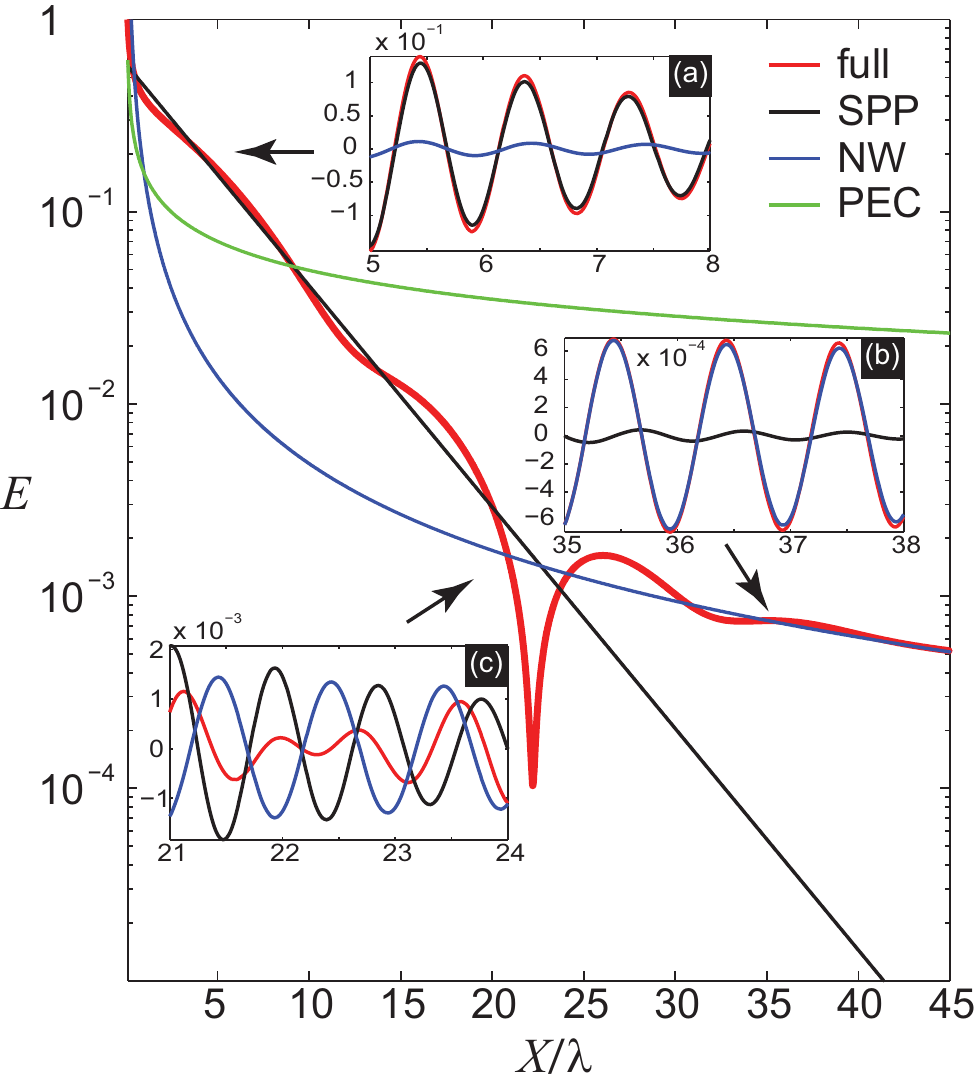}
\caption{(color online). The dependency upon distance from the slit of the electric field modulus at the gold surface. The wavelength is $\lambda=540$ nm and $q_y=0$.
Together with the total field, both SPP and NW contributions, and the PEC case are also plotted. The insets show the real part of the $z$-component of the total, SPP and NW fields, in different spatial regions. All fields are normalized to $E(X=0.1\lambda,z=0)$, which is taken as a representative value in the near field.}\label{Fig3}
\end{figure}

Before presenting the dependence of the field pattern on $q_y$, for completeness and in order to make the comparative analysis easier, we briefly review the case $q_y=0$ (extensively studied in ~\onlinecite{SoukoulisPRB09,NikitinNJP09,NikitinPSS10,LalanneSciRep09}).
Figure~\ref{Fig3} shows a representative case ($\lambda=540$ nm and $q_y=0$) for the dependence on distance to the slit of the electric field at the metal surface.
The fields for both vacuum-gold and vacuum-PEC interfaces are shown, under the assumption that they are launched by slits with the same amplitude of the electric field on their exit apertures. The spatial dependence of the field is quite different in these two cases. Instead of the cylindrical wave-type algebraic decay $\propto1/\sqrt{x}$ along the PEC surface, the field along the vacuum-gold interface shows two different behaviors, separated at the crossover distance, $X_c$. It must be stressed that the chosen wavelength, $\lambda=540$ nm, and the type of the metal (gold) does not represent a special case with some particular properties, but corresponds to a small value of $X_c$. Otherwise, the physics for this case is as rich as for other wavelengths.

As shown in Fig.~\ref{Fig3}, in a very close vicinity to the slit $x\ll1$, the behavior of the field is complex and contains the contribution from all the angular spectrum of the Green's function, or in other words, from all the density of EM states. Phenomenologically, the field in this region has been represented by a SPP plus an additional contribution (defined as the total field minus the SPP one) denoted either ``creeping wave'' or ``quasi-cylindrical wave'' (CW, see Refs.~\onlinecite{LalanneNature06,LalanneSciRep09}). As mentioned before,  Eq.~\eqref{e5} faithfully represents the field (and thus the CW) for $X\geq0.1\lambda$. As $X$ increases, all the smooth parts in the angular spectra in the integral are progressively canceled out due to integration with the oscillatory factor $\sim\exp(iq_xx)$, so that only the sharp regions (with width $\Delta q \sim 1/x$) of the spectrum give a finite net contribution. These regions correspond to the vicinities of either the pole $q = q_p$ (which is a feature of a finite width) or the kink $q=1$ (which has zero width in $q$-space). The field in the region where these two contributions dominate, can be found by asymptotically expanding Eq.~\eqref{e5}, see Subsection~\ref{Sff}.

The electric field corresponding to the SPP and NW terms are rendered in Fig.~\ref{Fig3}. At distances from the slit of order of one wavelength, both field amplitude and phase [see inset (a) in Fig.~\ref{Fig3}] are well approximated by the SPP contribution, and the influence of the NW is weak. The field is locally enhanced comparing to the PEC case, and the efficiency of the SPP excitation depends upon wavelength. In the region close to the asymptote in the SPP dispersion relation ($\varepsilon_m\simeq-1$), the density of electromagnetic states increases and so it does the local field enhancement. However, while the mode becomes both slower and more confined, due to the increase of its wavevector, the absorption increases as well and the SPP mode is quenched at a smaller distance from the source.

\begin{figure}[thb!]
\includegraphics[width=8cm]{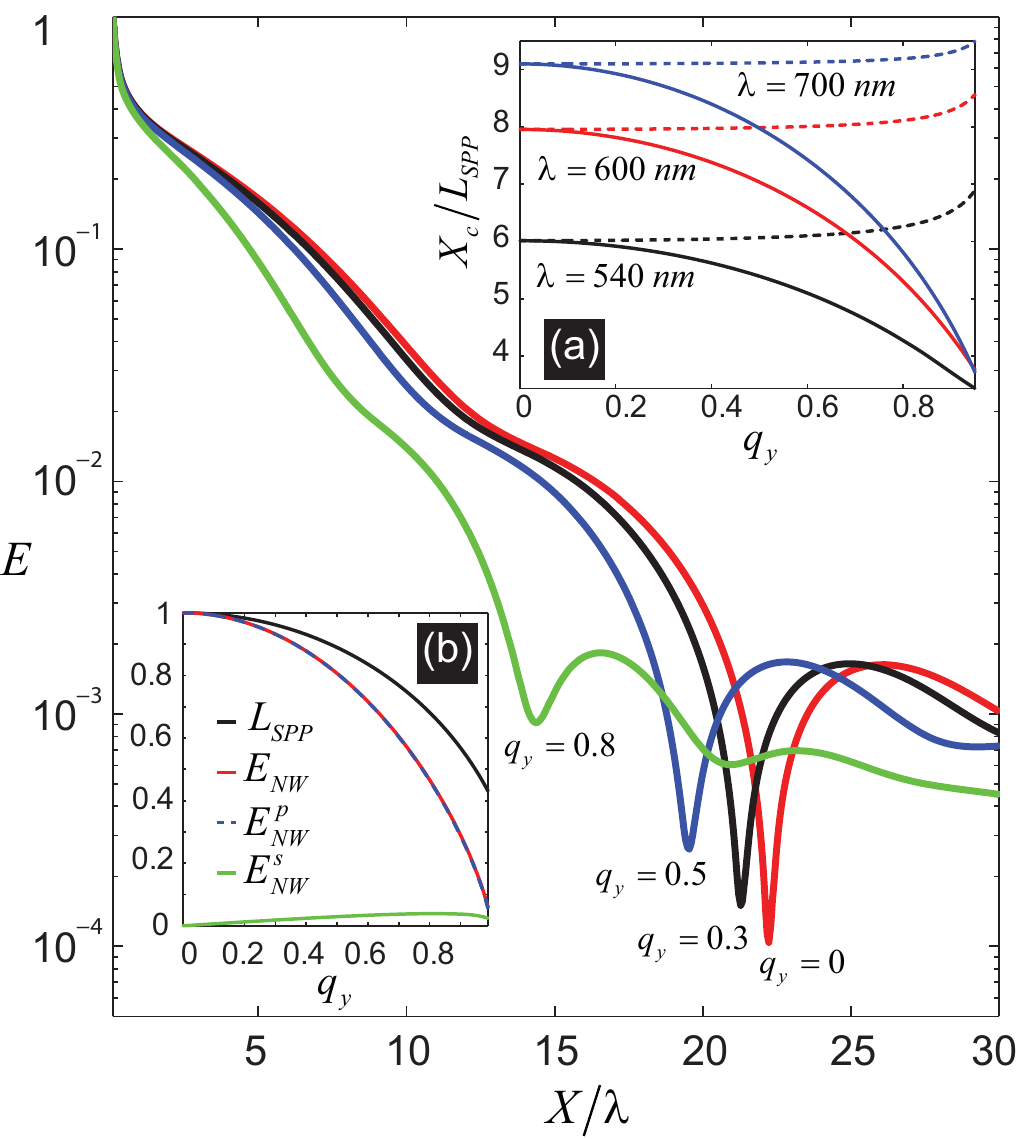}
\caption{(color online). The dependency on distance from the slit of the electric field modulus, at the vacuum-gold surface for different values of $q_y$. The wavelength is $\lambda=540$ nm. The normalization of the fields is the same as in Fig.~\ref{Fig3}. The inset (a) shows the dependency of the ratio between the crossover distance and SPP propagation length ratio, $X_c/L_{SPP}$, as a function of $q_y$ and for different wavelengths. The continuous curves are for $X_c(q_y)/L_{SPP}(q_y=0)$, while the discontinuous ones are for $X_c(q_y)/L_{SPP}(q_y)$. The inset (b) renders the normalized SPP propagation length, $L_{SPP}(q_y)/L_{SPP}(q_y=0)$ and the NW field modulus $E_{NW}$, together with its $p$- and $s$-polarization components, all of them normalized to $E_{NW}(q_y=0)$.}\label{Fig4}
\end{figure}

At distances large enough so that, due to the absorption, the SPP is sufficiently damped, the contributions from SPP and NW are comparable, see inset (b) in Fig.~\ref{Fig3}. This typically occurs at $X_c\sim6-9 L_{SPP}$, with $L_{SPP}$ being the SPP propagation length. In the vicinity of $X_c$, the SPP and NW fields have similar amplitudes, so the modulus of their sum presents an interference behavior, leading to  a set of maxima and minima. Notice that, in the optical region the SPP wavevector is close to the light cone, thus close to the NW one. However, $q_p$ largely increases close frequencies such that $\varepsilon_m\simeq-1$ (which, for good metals, occur at $\omega_p/\sqrt{2}$, where $\omega_p$ is the plasma frequency), in which case the total field given by Eq.~\eqref{e9} presents a fanciful two-scaled oscillatory behavior.

The third region is located beyond $X_c$, where the contribution from the SPP field is negligible (see inset (c) in Fig.~\ref{Fig3}). The field then reaches its asymptotic behavior, which is given by the NW, oscillating with the spatial period given by the free-space wavelength and decaying algebraically as $\sim1/x^{3/2}$.

Let us now analyze the dependency of the fields upon the $y$-component of the wavevector. Fig.~\eqref{Fig4} shows the spatial dependencies of the fields in the direction perpendicular to the slit for different $q_y$. There are two tendencies with the increase of $q_y$: the amplitude of the field decreases, and the crossover distance $X_c$ diminishes. To explain this behavior, we represent $X_c$ as a function of $q_y$ (inset (a) to Fig.~\ref{Fig4}). As can be seen from the curves where $X_c$ is normalized to the constant ($q-$independent) value $L_{SPP}(q_y=0)$, the distance after which the NW dominates decreases as $q_y$ increases.
However, the value of $X_c(q_y)/L_{SPP}(q_y)$ increases as $q_y$ increases, meaning that the crossover occurs at smaller absolute distances, but at larger SPP propagation lengths, specially close to $q_y=1$.

The inset (b) of Fig.~\ref{Fig3} shows the NW amplitude as a function of $q_y$.
This behavior is due to the dependence with $q_y$ of the space-independent prefactors in the amplitude of the NW (which for $|\varepsilon_m|\gg1$, goes as $E^p_{NW}\propto q_0^{3/2}$, and thus decreases as $q_y$ increases). By contrast, in the case of a SPP this prefactor is practically independent of $q_y$. However, the SPP propagation length scales as
$L_{SPP}\propto q_0$, so the SPP decays faster for larger values of $q_y$.
As a result, as $q_y$ increases, the NW overtakes the SPP closer to the slit, but with a smaller amplitude.

As follows from Eqs.\eqref{e7},\eqref{e11}, the $s$-polarization component of the NW has a non-monotonic dependency upon $q_y$, $E^s_{NW}\propto q_y\sqrt{q_0}=q_y \sqrt{\varepsilon-q_y^2}$. Nonetheless, the contribution of this component is always much smaller than that from $p$-polarization (see inset (b) of Fig.~\ref{Fig4}). Notice that, as the NW presents the same algebraic decay for all $q_y$ ($E_{NW}\propto 1/x^{3/2}$), the normalization to $E_{NW}(q_y=0)$ makes the quantities represented in the inset (b) of Fig.~\ref{Fig4} independent upon distance.
Thus, even for oblique incidence, NWs are virtually $p$-polarized waves along the interface, i.e., present the same polarization as SPPs.

\begin{figure}[thb!]
\includegraphics[width=8cm]{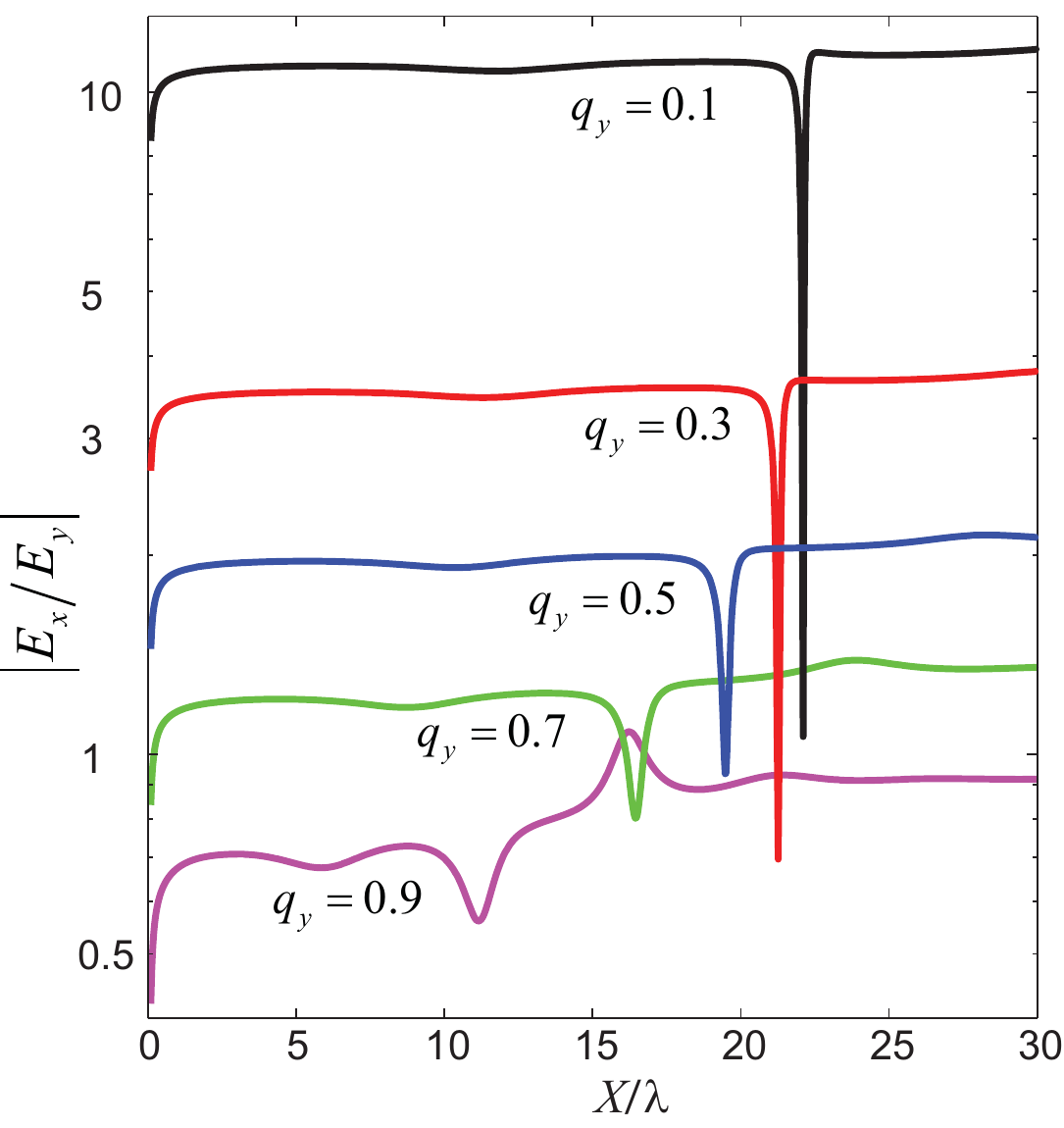}
\caption{(color online). The dependency of the ratio $E_x/E_y$ upon distance from the slit for different values of $q_y$. The wavelength is $\lambda=540$ nm and the considered metal is gold.}\label{Fig5}
\end{figure}

It is also interesting to study the ratio $E_x/E_y$, i.e. the polarization of the tangential-to-the-interface component of the field. For the case of in-plane launching ($q_y=0$), $E_y=0$, and the field in the plane of the metal surface has only $x$-component. When the excited waveguide mode gains non-zero momentum in the direction along slit, the scattered electric field possesses a finite $y$-component in both far- and near-field regions. Fig.~\ref{Fig5} illustrates the spatial dependency of $|E_x/E_y|$ for different $q_y$, showing that this ratio increases with $q_y$. According to Eqs.~\eqref{e6}, for a SPP the ratio $E_x/E_y$ scales as $\sqrt{(q_p/q_y)^2-1}$.  For a NW, from Eq.~\eqref{e7} and neglecting the contribution from the $s$-polarization (which scales as $O(1/\varepsilon_m)$), it follows that this ratio scales in a similar way:  $E_x/E_y\simeq\sqrt{1/q_y^2-1}$. Notice  that crossover distance is slightly different for $x$ and $y$-components of the electric field  and, also, that the amplitudes of $E_x$ and $E_y$ are different at the crossover. For these reasons, the curves in Fig.~\ref{Fig5} present dips at distances close to the corresponding crossovers, where both $E_x$ and $E_y$ are strongly suppressed, but $E_y$ dominates.

\section{Conclusions}

To conclude, in this paper we have presented the asymptotic expression for the EM field along the metal interface, launched by a subwavelength slit in the conical mount. These expressions are very accurate for the region down to a tenth of a wavelength. The field presents different contributions, which can be assigned to specific regions in the angular spectrum. The $s$-polarization component, although present for oblique incidence, not influence significantly the behavior of the field. We have studied the dependencies of the fields upon angle of incidence (i.e., component of the EM wavevector along the slit axis). In the far field region, the distance from the slit at which the algebraic behavior of the field overcomes the exponential decay decreases as $q_y$ increases. This could favor experimental studies of the Norton waves.

\section{Acknowledgements}

The authors acknowledge support from the Spanish Ministry of Science and
Innovation under grants MAT2009-06609-C02 and CSD2007-046-NanoLight.es. A.Y.N.
acknowledges the Juan de la Cierva grant JCI-2008-3123.

\appendix*

\section{\label{App}Mathematical treatment of the field and Green's Dyadic}

In this Appendix we present the details of the analytical computations for obtaining the field along the metal interface. The coordinate system is shown in Fig.~\ref{Fig1} with $Z=0$ corresponding to the exit interface.

The general self-consistent form of the field is given by the Lippmann-Schwinger integral equation\cite{OMartinPRL95}
\begin{equation}\label{a1}
\begin{split}
\mathbf{E}(\mathbf{R}) = \mathbf{E}_0(\mathbf{R}) + k_\omega^2\int_Vd\mathbf{R}' \Delta\varepsilon(\mathbf{R}') \hat{\mathcal{G}}(\mathbf{R},\mathbf{R}')\mathbf{E}(\mathbf{R}'),
\end{split}
\end{equation}
where $\mathbf{E}_0$ is the solution without the slit, $\Delta\varepsilon(\mathbf{R})=\varepsilon-\varepsilon_m$ is the variation of the dielectric permittivity in the volume occupied by the slit $V$ ($\Delta\varepsilon=0$ outside the slit).
The film is supposed to be optically thick, so that $\mathbf{E}_0 =0$ in the region of transmission, and the Green's function dyadic $\hat{\mathcal{G}}$ can be approximated by the one for the metal-vacuum interface, which satisfies the
equation
\begin{equation}\label{a1.1}
\begin{split}
\nabla \times\nabla \times \hat{\mathcal{G}}(\mathbf{R},\mathbf{R}') - k_\omega^2\epsilon \hat{\mathcal{G}}(\mathbf{R},\mathbf{R}') = \hat{I}\delta(\mathbf{R}-\mathbf{R}') ,
  \end{split}
\end{equation}
with standard boundary conditions at $z=0$. In Eq.~\eqref{a1.1} $\hat{I}$ is a diagonal unit matrix; $\epsilon=\varepsilon$ inside the dielectric and $\epsilon=\varepsilon_m$ inside the metal. For convenience, let us introduce the dimensionless coordinates $x,y,z=k_\omega X,k_\omega Y,k_\omega Z$. Then we assume that the slit is thin enough so that the field inside it has predominantly $x$-component and therefore only the $x$-column of $\hat{\mathcal{G}}$ will be essential, $\boldit{\mathcal{G}}=\mathbf{e}_x\hat{\mathcal{G}}$. This vector is represented by a two-dimensional integral in $k$-space (see Ref.~\onlinecite{Novotny}), and can be decomposed in $p$- and $s$-polarization contributions $\boldit{\mathcal{G}}=\boldit{\mathcal{G}}^{p}+\boldit{\mathcal{G}}^{s}$, where
\begin{equation}\label{a2}
\begin{split}
\boldit{\mathcal{G}}^{s,p}(\mathbf{r},\mathbf{r}') = \frac{ik_\omega}{8\pi^2}\int dq_xdq_y \mathbf{a}^{s,p} e^{i(\mathbf{q}\boldit{\rho}-q_{zm}z'+q_{z}z)},\\
\end{split}
\end{equation}
$\mathbf{q}=(q_x,q_y)$ with $q_{x,y}=k_{x,y}/k_\omega$, $q_z=\sqrt{\varepsilon-q^2}$, $q_{zm}=\sqrt{\varepsilon_m-q^2}$ and $\boldit{\rho}=(x-x',y-y')$. The vectors $\mathbf{a}^s$, $\mathbf{a}^p$ are defined to be
\begin{equation}\label{a3}
\begin{split}
\mathbf{a}^p = \frac{t^p}{q^2\sqrt{\varepsilon_m\varepsilon}}
\begin{pmatrix}
q^2_xq_z\\
q_xq_yq_z\\
-q_xq^2
\end{pmatrix},\q
\mathbf{a}^s = \frac{t^s}{q^2q_{zm}}
\begin{pmatrix}
q^2_y\\
-q_xq_y\\
0
\end{pmatrix},
\end{split}
\end{equation}
with $t^p$, $t^s$ being the Fresnel coefficients given by Eq.~\eqref{e3}. Taking into account that the field inside the slit is given by the fundamental mode and taking into account the momentum conservation along $y$, we can write
$\mathbf{E}(x',y',z') = \mathbf{e}_xE(x',z')e^{iq_{y0}y'}$,
where $q_{y0}$ is the dimensionless $y$-component of the incident wave wavevector. Extracting the $y$-dependency of the Green's dyadic, we have the following integral
\begin{equation}\label{a4}
\begin{split}
\int\limits_{-\infty}^\infty dy' e^{i(q_{y0}-q_y)y' + iq_{y}y} = 2\pi \delta(q_{y0}-q_y)e^{iq_{y0}y}.
\end{split}
\end{equation}
Then the integration in Eq.~\eqref{a2} in $q_y$ is performed trivially. From Eq.~\eqref{a2} it follows that the integrand contains the exponential factor $e^{-iq_{zm}z'}$, which decays at the distance of a skin
depth $\delta = 1/\mathrm{Im}(q_{zm})$, and is of the order of a few tens of nm in the optical regime. We can, therefore,
extend the integration limits in $z'$ to $[-\infty, 0]$. Additionally, the variation of the
dyadic is much faster than that of the field inside the slit, hence the electric field inside
the slit can be approximated by its value at the distance $z = -\delta$ (the average
distance to the surface, weighted by the exponential decay of the field)
\begin{equation}\label{a5}
\begin{split}
\int\limits_{-h}^0 dz' e^{-iq_{zm}z'}\simeq \int\limits_{-\infty}^0 dz' e^{-iq_{zm}z'} = \frac{i}{q_{zm}}\simeq\frac{i}{\sqrt{\varepsilon_m-\varepsilon}}.
\end{split}
\end{equation}
Then Eq.~\eqref{a1} becomes
\begin{equation}\label{a6}
\begin{split}
\mathbf{E}(x,y,z) = C \int_L dx'\mathbf{G}(x-x',y,z)E_x(x',z'=-\delta),
\end{split}
\end{equation}
where $C = i\sqrt{\varepsilon_m-\varepsilon}/k_\omega$ and $\mathbf{G}=\mathbf{G}^{p}+\mathbf{G}^{s}$ with
\begin{equation}\label{a7}
\begin{split}
\mathbf{G}^{s,p}(x-x',y,z) = \frac{ik_\omega}{4\pi}\int dq_x \mathbf{a}^{s,p} e^{i[q_x(x-x')+ q_yy+q_{z}z]}.
\end{split}
\end{equation}
For brevity we have omitted ``0'' in $q_{y0}$. We have thus recovered Eqs.~\eqref{e1}-\eqref{e3}.

Taking into account the presence of the poles (placed at $q_{z}\varepsilon_m + q_{zm}\varepsilon=0$ ) and branch cuts and branch points (defined by $\mathrm{Im}(q_{z})=0$), an asymptotic analysis of the integral \eqref{a7} can be made following the general recipes described Ref.~\onlinecite{FelsenMarcuvitz}, as was done for the case $q_y=0$ in Ref.~\onlinecite{NikitinNJP09}. Concentrating on the field at the interface $z=0$, we simplify $\mathbf{G}$ in the following way
\begin{equation}\label{a8}
\begin{split}
\mathbf{G}(x,z=0) \equiv \frac{ik_\omega}{2\pi}\mathbf{g}(x).
\end{split}
\end{equation}
The branch cuts $\mathrm{Im}(q_z)=0$ can be removed by changing to polar variables: $q_z =q_0\cos\phi$, or $q_x=q_0\sin\phi$ with $q_0^2=\varepsilon-q_y^2$. Here we assume that the contribution of the branch cuts $\mathrm{Im}(q_{zm})=0$ is negligible, being of order $\sim e^{-|\sqrt{\varepsilon_m}|x}$ (otherwise, some modifications in the solution scheme would be necessary). Then, to provide the exponential decay of the integrand, the variable $\phi$ is further transformed into the variable $s$ as follows: $\sin\phi=1+is^2$, so that the saddle point is placed at $s=0$.  With this change the vector $\mathbf{g}$ reads
\begin{equation}\label{a9}
\begin{split}
&\mathbf{g}(x) = e^{iq_0x}\int_L ds \boldit{\Phi}(s)e^{-q_0xs^2}, \\
&\boldit{\Phi}(s) = \frac{d\phi}{ds}\cdot \frac{\mathbf{a}[q_x(s)]}{2}q_0\cos[\phi(s)],
\end{split}
\end{equation}
where the integration path $L$ corresponds to the real axis in the complex plane $q$. Then the singular part of the integrand is separated
\begin{equation}\label{a10}
\begin{split}
\boldit{\Phi}(s)=\frac{\mathbf{C}_p}{s-s_p} + \boldit{\Phi}_0(s), \; \boldit{\Phi}_0(s)=\frac{\boldit{\Phi}(s)(s-s_p)-\mathbf{C}_p}{s-s_p},
\end{split}
\end{equation}
where $s_p = e^{-i\frac{\pi}{4}}\sqrt{q_{xp}/q_0-1}$; $q_{xp}$ being the $x$-component of the SPP wavevector, which satisfies, $q_p^2=q_{xp}^2+q_y^2=\varepsilon\varepsilon_m/(\varepsilon+\varepsilon_m)$. We would like to remind that only $p$-polarization type of poles are physical (i.e. satisfy the radiation condition $\mathrm{Im}(q_{z})>0$), and therefore the separation \eqref{a10} has sense only for $p$-polarization component of $\boldit{\Phi}$. The elements of the vector $\mathbf{C}_p$ are given by the residues of the integrand defined by Eq.~\eqref{e6}.

Transforming the integration contour $L$ to the real axis in the complex plane $s$, the singular part of $\mathbf{g}$ can be represented using the complementary error function $\mathrm{erfc}$
\begin{equation}\label{a11}
\begin{split}
&\mathbf{g}(x) = i\pi\, \mathbf{C}_pe^{q_0x(i-s_p^2)}\, \mathrm{erfc}(- is_p\sqrt{q_0x}) +
 \mathbf{g}_0(x), \\
&\mathbf{g}_0(x) = e^{iq_0x}\int\limits_{-\infty}^{\infty} ds\, e^{-q_0x s^2}\boldit{\Phi}_0(s).
\end{split}
\end{equation}
The integral appearing in the nonsingular term of $\mathbf{g}$ can be represented in the form of an infinite sum resulting from the integration of the Tailor expansion for $\boldit{\Phi}_0(s)$. This series reads
\begin{equation}\label{a12}
\begin{split}
\mathbf{g}_0(x) = e^{iq_0x}\sum\limits_{n\in\mathrm{even}}\frac{1}{n!}\frac{d^n\boldit{\Phi}_0(s)}{ds^n}|_{s=0}\frac{\Gamma(\frac{1+n}{2})}{(q_0x)^\frac{1+n}{2}},
\end{split}
\end{equation}
where $\Gamma$ is Gamma function. For a large range of $x$ only the two first terms of this expansion are important. With these two terms in Eq.~\eqref{a12}, i.e., with the precision up to $O(x^{5/2})$, Eq.~\eqref{a11} transforms to Eqs.~\eqref{e5}-\eqref{e7}.

In the far-field region, which formally occurs for $|s_p|\sqrt{q_0x}\gg1$ (although comparisons with the numerical computation of the field shows that, in practice, this ``far-field'') the asymptotic expansion of the complementary error function can be used:
\begin{equation}\label{a13}
\begin{split}
\mathrm{erfc}(-is_p \sqrt{q_0x})= 2 + \frac{e^{s_p^2 q_0x}}{s_p \sqrt{\pi q_0x}}\sum\limits_{n=0}^\infty\frac{(-1)^n}{(-i)^{2n+1}}\frac{(2n)!}{n!(4s^2_p q_0x)^n},
\end{split}
\end{equation}
where we have taken into account that $\mathrm{Im}(s_p)<0$. Formally, this asymptotic expansion is valid for distances such that $|s_p|\sqrt{q_0x}\gg1$. However, comparisons with the exact results for the fields (obtained form the numerical computation of relevant integrals)  shows that usually this condition is too restrictive and the asymptotic expansion is valid even for shorter distances.

\end{document}